\documentclass[preprint,prb,tighten,floatfix,showpacs]{revtex4}
\usepackage{graphicx}

\def\<{\langle}
\def\>{\rangle}

\def\be{\begin{equation}}
\def\ee{\end{equation}}

\begin{document}
\preprint{cond-mat} \title{ Entanglement Entropy of Quantum Hall Systems at Half Filling }

\author{C. Balusek, B. A. Friedman, G. C. Levine* and D. Luna**}

\address{Department of Physics, Sam Houston State University, Huntsville TX 77341}

\address{*Department of Physics and Astronomy, Hofstra University,
Hempstead, NY 11549}

\address{**Department of Mathematics, Sam Houston State University, Huntsville 
TX 77341}

\date{\today}

\begin{abstract}    The entanglement entropy of $\nu=1/2$ and $\nu=9/2$ quantum Hall states in the presence of short range disorder has been calculated by direct diagonalization.  Spin polarized electrons are confined to a single Landau level and interact with long range Coulomb interaction.  For $\nu=1/2$ the entanglement entropy is a smooth monotonic function of disorder strength.  For $\nu=9/2$ the entanglement entropy is non monotonic suggestive of a  solid-liquid phase transition.  As a model of the transition at $\nu=1/2$ free fermions with disorder in 2 dimensions were studied.  Numerical evidence suggests the entanglement entropy scales as $L$ rather than the $L \ln{L}$ as in the disorder free case.  \end{abstract}

\pacs{03.67.Mn,73.43.Cd, 71.10.Pm}
\maketitle
\section{Introduction}

Entanglement entropy is a quantity, which, broadly speaking, quantifies the quantum mechanical entanglement between a system and a subsystem.  It may be useful as a numerical means to detect quantum phase transitions, in particular in cases for which an order parameter is non local or simply unknown \cite{zoz}.  For example, for the quantum Hall state at filling $\nu=5/2$, i.e. where the $N=1$ Landau level is half filled, there appears to be a  cusp, a lack of smoothness in the entanglement entropy, as a function of disorder strength \cite{FLL}.  This lack of smoothness, is apparent, in numerical calculations, even for small system sizes and is taken to be a signature of a quantum phase transition away from the Moore-Read state as disorder increases.

In this paper, the behavior of the entanglement entropy as a function of disorder strength is investigated for other half-filled systems, in particular for filling 1/2, the $N=0$ Landau level being 1/2 filled, and for filling 9/2, a 1/2 filled $N=2$ Landau level.  Since the physics of the $N=0,1,2$ Landau levels is quite different, quite distinct behavior as a function of disorder strength is anticipated.  In the lowest Landau level ($N=0$), the composite fermion picture provides a very good description of the essential physics.  In the half filled case each electron takes two flux quantum leaving composite fermions in two dimensions moving in zero magnetic field \cite{Jain}.  Although disorder can have drastic effects in two dimensions, due to the limited system sizes accessible with direct diagonalization, it is anticipated there is smooth behavior of the entanglement entropy as a function of disorder strength.  In the $N=2$, third Landau level, the electrons form stripes and bubbles or possibly liquid crystalline states \cite{Du,Lilly,Koulakov,Fradkin}.  In experiment, the stripes are very sensitive to disorder.  (More precisely, the anisotropy in the diagonal resistivity is sensitive to sample quality).  One therefore conjectures, in a numerical calculation one should see something analogous to a solid liquid transition, i.e. a signature of a first order phase transition.

\section{Entanglement Entropy as a Function of Disorder Strength for $\nu=1/2$ and $9/2$ }

The numerical method used is direct diagonalization applied to square (aspect ratio one) clusters with periodic boundary conditions, the square torus geometry.  As is standard, the Landau gauge is taken for the vector potential.  Spin-polarized electrons are confined to a single Landau level and interact with the Coulomb potential.  The random potential U(r) is taken to be delta correlated $< U(r)U(r')> = U_0 \delta (r-r')$ and the disorder strength is given by a parameter $U_R = \sqrt{3U_0/2}$.  This approach has previously been used to study the entanglement entropy without a disorder potential\cite{FL1} and with a disorder  potential at fillings $\nu=1/3$ and $\nu=5/2$ \cite{FLL}.  Due to lack of translational invariance and the necessity of disorder averaging  the largest system size considered was 13 electrons in 26 orbitals with a state space of approximately $1 X 10^7$.  Since in the third Landau level, the states are not necessarily liquid like, it would of interest to use clusters of different aspect ratios and pick the aspect ratio giving the lowest disorder averaged energy \cite{Shibata}.  However, because of computer time limitations this was not attempted.  

The entanglement entropy is calculated by taking a subsystem of $l$ adjacent orbitals. Recall in the Landau gauge, the orbitals consist of strips oriented along, say the y-axis, of width the order the magnetic length.  The reduced density matrix is straightforward to calculate from the ground state wave function.  It is then diagonalized, giving the eigenvalues $\lambda_j$ from which the $l$ orbital entanglement entropy $S(l)$ , $S(l)= -\sum_j{\lambda_j \ln{\lambda_j}}$ is obtained.  This procedure is done for every realization of the random potential; the results are averaged to give $<S(l)>$ where $< >$ denotes average over the random potential. For the smallest systems (10 electrons in 20 orbitals) we have averaged over 200 realizations of the random potential; the largest systems, we have averaged over as few as 30 realizations.  

Considering first, the lowest Landau level, in figure 1a $<S(2)>$ is plotted vs. disorder strength for filling 1/2.  $l=2$ corresponds to a real space width for the largest system size (N=26) of $2\sqrt{2\pi/26}$.  This is roughly one magnetic length.  The system sizes 10/20 (10 electrons in 20 orbitals), 11/22, 12/24 
and 13/26 were treated.  Error bars are given by the root mean square (rms) values of S(l), i.e. $\sigma= \frac{\sqrt{<S(l)^2>-<S(l)>^2}}{\sqrt{N_s-1}}$ with $N_s$ being the number of samples.  One sees rather smooth behavior vs. disorder strength in comparison to $\nu=5/2$, see fig 8 a of reference \cite{FLL}.  Similar behavior is obtained for larger subsystem sizes, in figure 1b $<S(12)>$ is plotted for $U_R$ for filling 1/2.  For the system sizes we could study numerically there is not much difference in considering a subsystem of fixed $l$ or a fixed aspect ratio subsystems.  That is, even quantitatively, there is not much difference in comparing $<S(12)> $ for all system sizes, as we have done,  or comparing $<S(10)>_{N=20}$, $<S(11)>_{N=22}$, $<S(12)>_{N=24}$, $<S(13)>_{N=26}$.
Once again, there is rather smooth behavior of the entanglement entropy vs. disorder strength.  This is at least consistent with what one would expect from non interacting composite fermions.  In two dimensions for non interacting fermions, all states are exponentially localized.  However, if the system size is smaller then the localization length, the states will appear to be extended  \cite{mac,kramer}.  We conjecture for the small system sizes  and disorder strengths we study, this is the case.  Experimentally \cite{wil} it appears the composite fermi liquid is more robust to the presence of disorder then the quantum Hall state at filling $5/2$.   
We also note that  qualitatively, the entanglement entropy vs. disorder curves for $\nu=1/2$ and $\nu=1/3$ are similar (see figure 7 of reference \cite{FLL}).       

Turning now to the third Landau level, figure 2a is a plot of $<S(2)>$ vs. disorder  and figure 2b is a plot of  $<S(12>$ vs. disorder for system sizes 10/20, 11/22,12/24 and 13/26.  For both $<S(2)>$ and $<S(12)>$  there is non monotonic behavior of the entanglement entropy as a function of $U_R$.  This is reminiscent of a first order phase transition (Maxwell construction) which, naturally, in the third Landau level can be interpreted as a transition from a solid to a liquid like state\cite{bernard}.  

\section{Entanglement Entropy of Free Fermions with Disorder at Half Filling}

We attempted for $\nu=1/2$ and fixed $l$ to perform a 1/N extrapolation (N is the system size).  Regrettably, this was not successful due to the nonlinear behavior with 1/N.  Note that in calculating the entanglement entropy for strips of $l$ orbitals (length $\sqrt{2 \pi N}$, width $l \sqrt{2 \pi /N}$ ), the subsystem, the strips, have a state space of dimension $2^l$.  Hence the reduced density matrix has a dimension $2^{l} X 2^{l}$ and the entanglement entropy is bounded from above by $l\ln{2}$ \cite{sch}.  It is thus impossible for the entanglement entropy to scale as the length of the strip, as $\sqrt{N}$ is greater then $l$ for large $N$.  By analogy to the area law for fermions \cite{gioev,wli,barthel}, we have investigated for disordered systems at 1/2 filling in the lowest Landau level, whether the entanglement entropy scales as $l^{1/2}$, $l$ or $l^{1/2} \ln{l}$.  We were unable to come to any definite conclusion due to the small system sizes.
 However, it seems, not without interest to study a simpler problem of the entanglement entropy of free fermions in zero magnetic field  at 1/2 filling with disorder in two dimensions.  The motivation for this model is the composite fermion picture where electrons in the 1/2 filled Landau level are equivalent to non interacting composite fermions in zero magnetic field.   That is, let us consider the usual Anderson tight binding Hamiltonian \cite{mac,kramer} in two dimensions on a square lattice with periodic boundary conditions.  There is nearest neighbor hopping $t$ with $t=1$ and a site diagonal matrix element that is rectangularly distributed with width W.  We take the system to be square, with periodic boundary conditions of dimension 60 X 60 and 80 X 80 and the subsystem to be small squares of dimension L X L where L ranges from 1 to 30.  The entanglement entropy was calculated with Peschel's method \cite{peschel} for a given realization of disorder and then an average over disorder is done giving $<S(L x L)>$.  Recall, even though the system is half filled, since the electrons are non interacting, this only involves one electron calculations and hence quite large systems and subsystems can be treated.  In figure 3a $<S(L x L)> $ is plotted vs. $L \ln{L}$ for W =15.  For this value of W, the localization length  is about 1 in units of the lattice constant (see the inset of fig. 1 of ref. \cite{mac}) and hence one anticipates both the system and subsystem to see the effect of localization.  An average over a 1000 realizations of the disorder was taken, the error bars  are barely discernible in the figure.  Recall, although all states are localized in 2 dimensions by disorder, W=15 was chosen to avoid complications due to finite size effects.  One notes a definite downward curvature in the graph 3a.  This is in contrast with  figure 3b in which $<S(L x L)>$ is plotted vs. L again for W=15 where one sees more linear behavior.  We suggest that  non interacting states with disorder  in dimension 2  obey an area law without multiplicative logarithmic corrections.  This result is a surprise in that 2 gapless fermi systems coupled by a weak link have an entropy proportional to $\ln{L}$ \cite{levine}  , hence multiple links should give a $L \ln{L}$ entropy.    In addition in one dimension, for  fermions, whether or not disorder is present, the entanglement entropy is proportional to $ \ln{L}$\cite{ref,la} .

\section{Remarks on the Topological Entanglement Entropy for $\nu=5/2$ Without Disorder}

Recently, it has become possible to compute the entanglement entropy for 18 electrons in 36 orbitals, with no disorder, in the flat torus geometry, without extreme numerical effort.  This is due to improvements in hardware; workstations with 64 GB of RAM are not very costly.  This technological improvement has enabled us to reconsider the following puzzling situation.  If system sizes 24, 26,28,30 and 32 are used to extrapolate the entanglement entropy S(l) and then the topological entanglement entropy $\gamma$ is extracted from the S(l) vs. $\sqrt{l}$ line, a value of $ \gamma = 2.07 \pm 0.15$ is obtained.  However, if the same procedure is followed including size 34, a "worse" value is found $\gamma = 1.93 \pm 0.16$ \cite{FL} in comparison to the value $\gamma =2.08$ for the Moore-Read state\cite{mr,nayak}.  In figure 4a, the extrapolated entanglement entropy using system sizes 12-18 is plotted vs. $l^{1/2}$.  The error bars in figure 4a correspond to an estimate of an error in the linear extrapolation.  The value of the y-intercept, $-\gamma = -2.08 \pm 0.19$.  Thus by including system size 36 much closer agreement with the Moore-Read value is obtained.  What is the origin of this behavior?  In figure 4b, $S(l)$ is plotted vs. $l^{1/2}$ for system sizes 32, 34 and 36.  It is hard to see system size 34 in that it is covered up by the system size 32 points.  The growth in the entanglement entropy occurs only when a pair of electrons is added.  This numerical effect is consistent with the Moore-Read state being a paired state of fermions.  Returning to figure 4a, the extrapolated entanglement entropies obtained from even number of electrons (system size 24, 28,32,36) is plotted as diamonds.  Fitting these points to a line, one obtains a y-intercept of $-\gamma =-2.20 \pm0.19$ again consistent with the topological entanglement entropy of the Moore-Read state.  These results agree with the analysis using the entanglement spectrum \cite{li,thom,zhao} applied to the quantum Hall spherical geometry, that the essential physics of $\nu=5/2$ is given by the Moore-Read state.

Finally, it was possible to obtain one more system size for 7/3 filling, 13 electrons in 39 states.  In figure 5, $S(l)$ is plotted vs. $l^{1/2}$ for various system sizes.  There is little change in the entanglement entropy is going from 8 to 9 electrons and 12 to 13 electrons.  Since 4 divides both 8 and 12, this behavior is consistent with the 4 fermion clusters in the k=4 Read-Rezayi\cite{rr} state.  This result, though consistent with our previous work \cite{FL}, is in disagreement with recent experiments\cite{dolev} that show that $\nu=7/3$ does not appear to be the k=4 Read-Rezayi state.  Note however, that our calculations do  not include disorder; we expect a state with 4 fermion clusters to be very sensitive to disorder given the sensitivity of the Moore-Read state to disorder \cite{FLL,gamez}.   

\section{Conclusions}

The entanglement entropy of $\nu=1/2$ and $\nu=9/2$ quantum Hall states in the presence of short range disorder has been calculated by direct diagonalization.  For $\nu=1/2$ the entanglement entropy is a smooth monotonic function of disorder strength.  For $\nu=9/2$ the entanglement entropy is non monotonic suggestive of a  solid-liquid phase transition.  As a model of the transition at $\nu=1/2$ free fermions with disorder in 2 dimensions were studied.  Numerical evidence suggests the entanglement entropy scales as $L$ rather than the $L \ln{L}$ as in the disorder free case.  It therefore may be of some interest to use the entanglement entropy as a numerical tool to study interacting electrons in zero magnetic field but in the presence of disorder, particularly in two dimensions.

We thank Mahito Kohmoto for a  helpful conversation.  This work was supported  in part by NSF Grant no. 0705048 (B.F. and D. L.) and the Department of Energy, DE-FG02-08ER64623---Hofstra University Center for Condensed Matter (G. L.).

\bigskip
\bigskip

\noindent
{\bf Figure Captions}
\bigskip

\noindent
figure 1. $<S(l)>$ vs. $U_R$ for filling $1/2$.  In figure 1a  $<S(2)>$ is  plotted, while figure 1b is a plot of  $<S(12)>$.
\bigskip

\noindent
figure 2. $<S(l)>$ vs. $U_R$ for filling $9/2$.  In figure 2a  $<S(2)>$ is  plotted, while figure 2b is a plot of  $<S(12)>$.
\bigskip

\noindent
figure 3.  $<S(L x L)>$  for an Anderson model on a two dimensional square lattice at 1/2 filling.  W = 15 and the system sizes are 60 x 60 and 80 x 80 with periodic boundary conditions.  In figure 3a $S(L x L)$ vs. $L\ln{L}$ is plotted.  In figure 3b $S(L x L)$ vs. $L$ is plotted.  An average over a 1000 realizations of the disorder was taken, the error bars  are barely discernible in the figures.
\bigskip

\noindent
figure 4. Entanglement entropy vs. $l^{1/2}$ for $\nu=5/2$, no disorder.  In figure 4a the extrapolated entanglement entropy is plotted.  In figure 4b the entanglement entropy for system sizes 16/32, 17/34 and 18/36 is plotted.  The error bars in figure 4a correspond to an estimate of an error in the linear extrapolation.
\bigskip

\noindent
figure 5.  $S(l)$ vs. $l^{1/2}$ for various system sizes $\nu=7/3$.

\eject

\begin{figure}[ht]
\includegraphics[height=22cm]{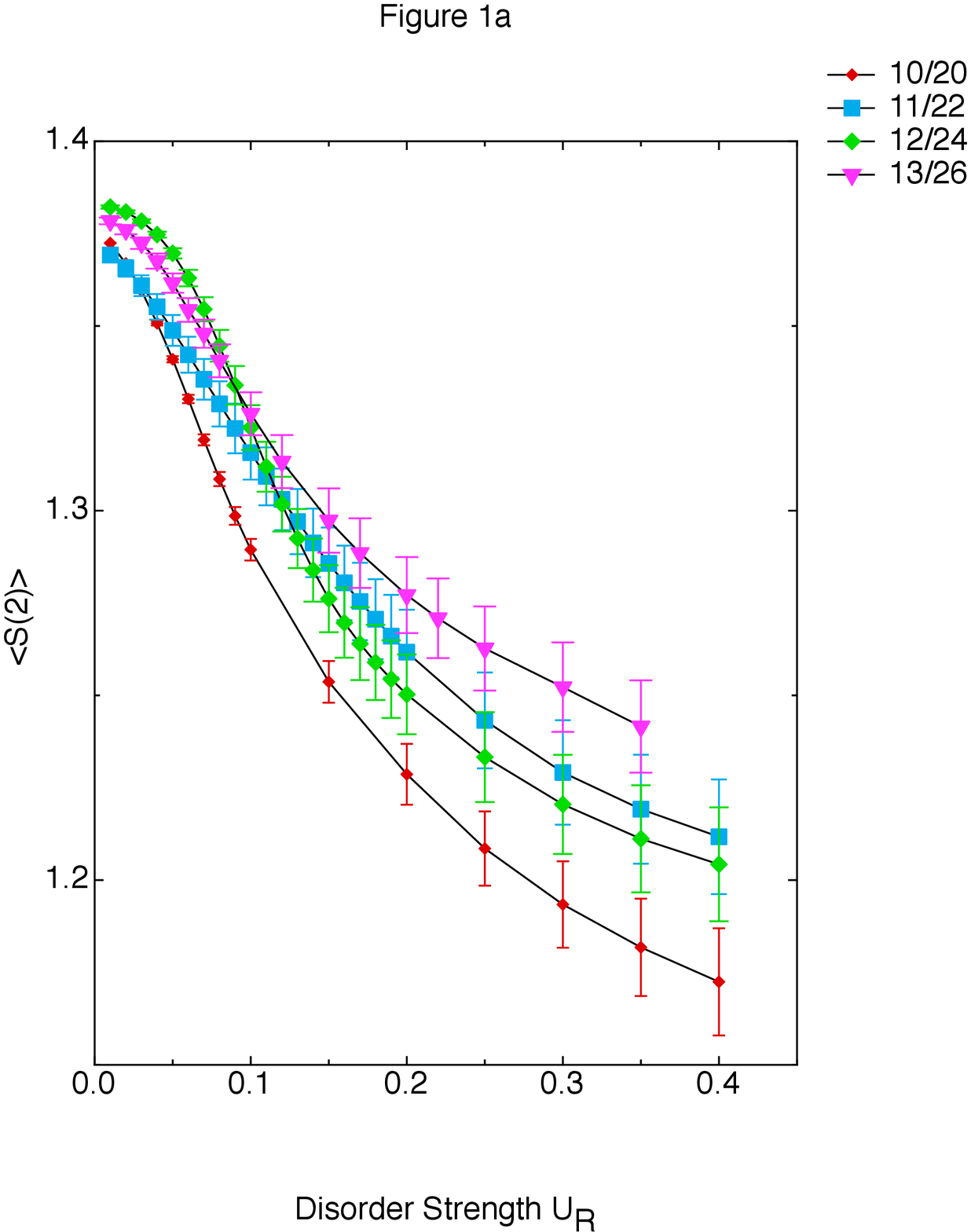}
\end{figure}

\begin{figure}[ht]
\includegraphics[height=22cm]{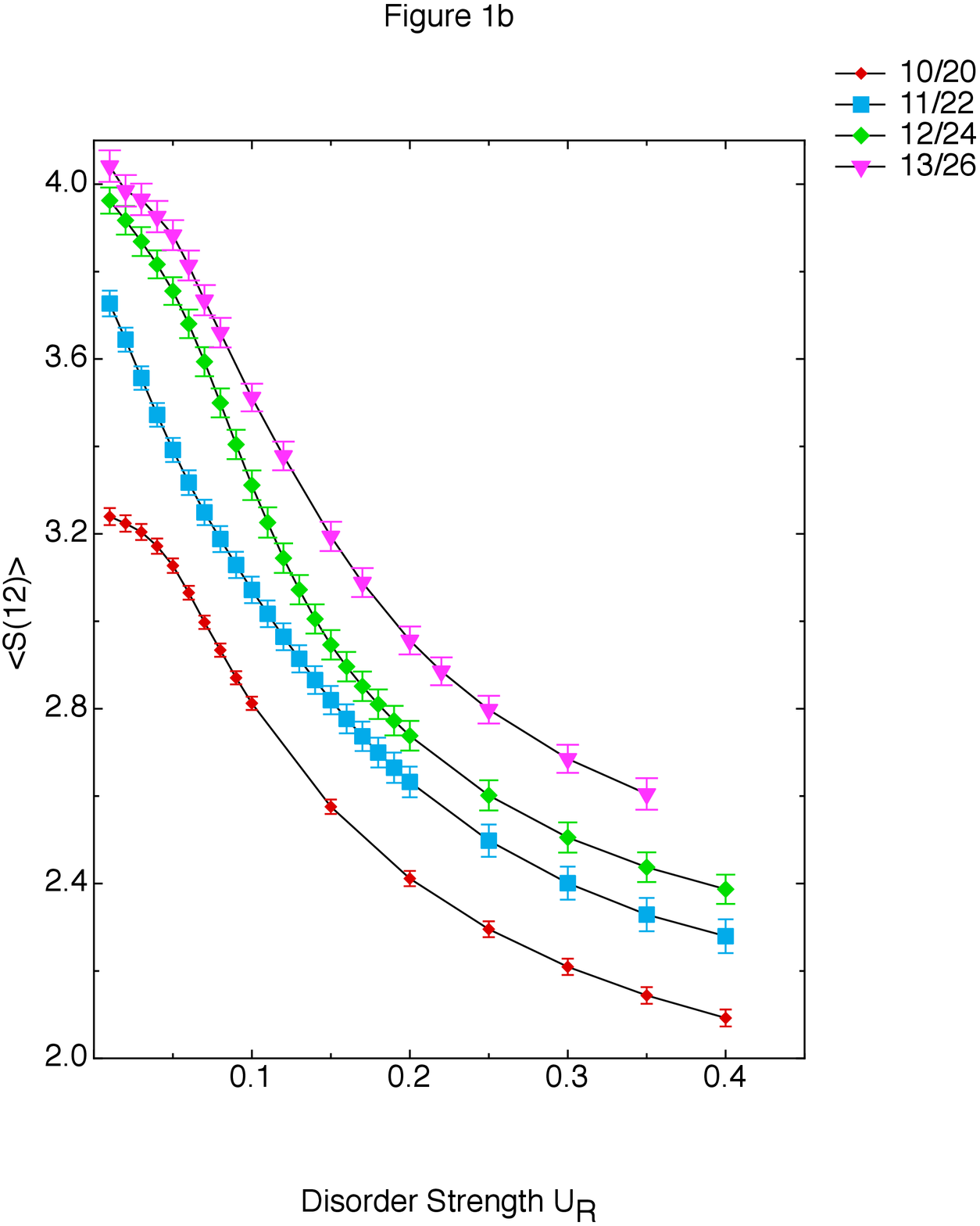}
\end{figure}

\begin{figure}[ht]
\includegraphics[height=22cm]{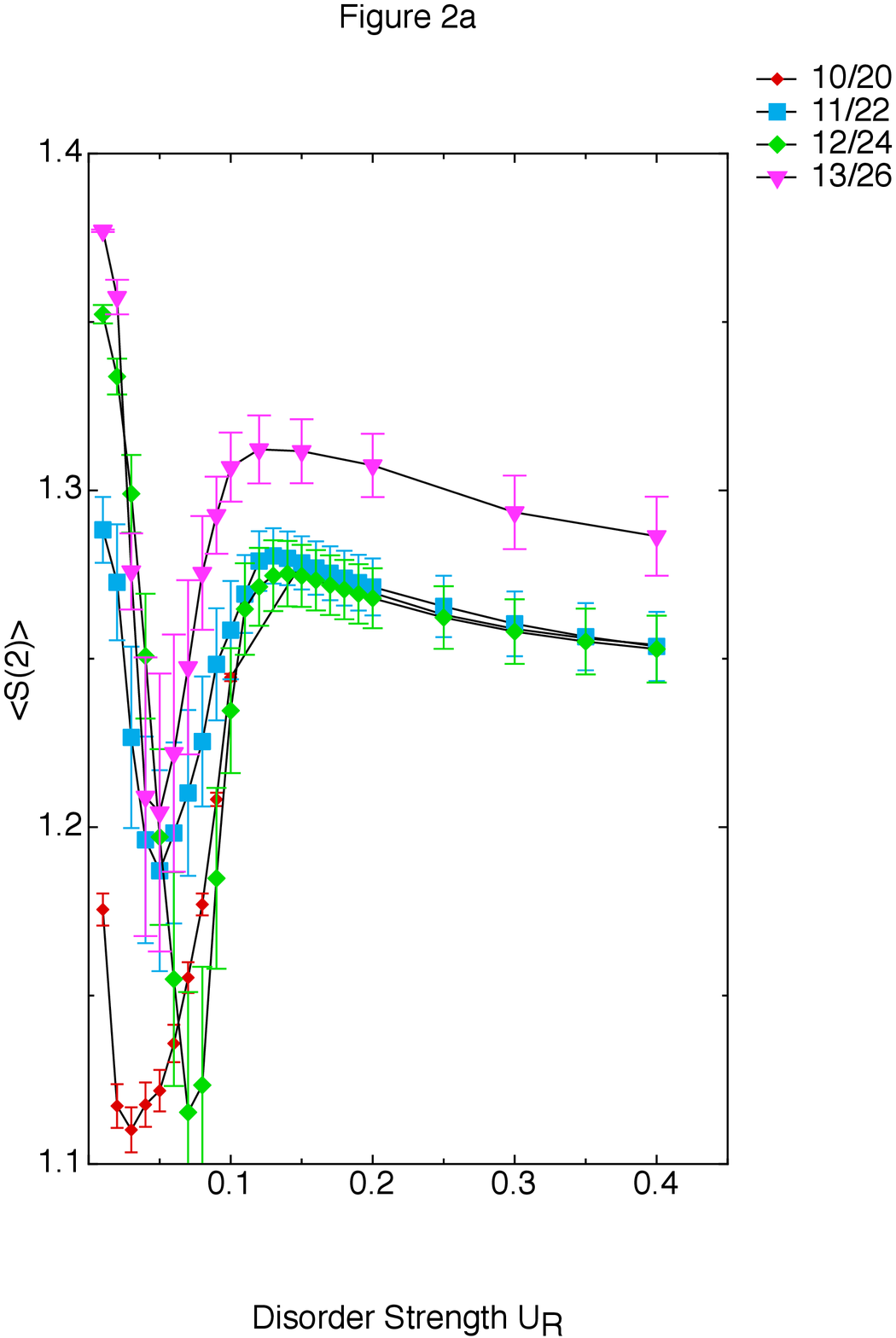}
\end{figure}

\begin{figure}[ht]
\includegraphics[height=22cm]{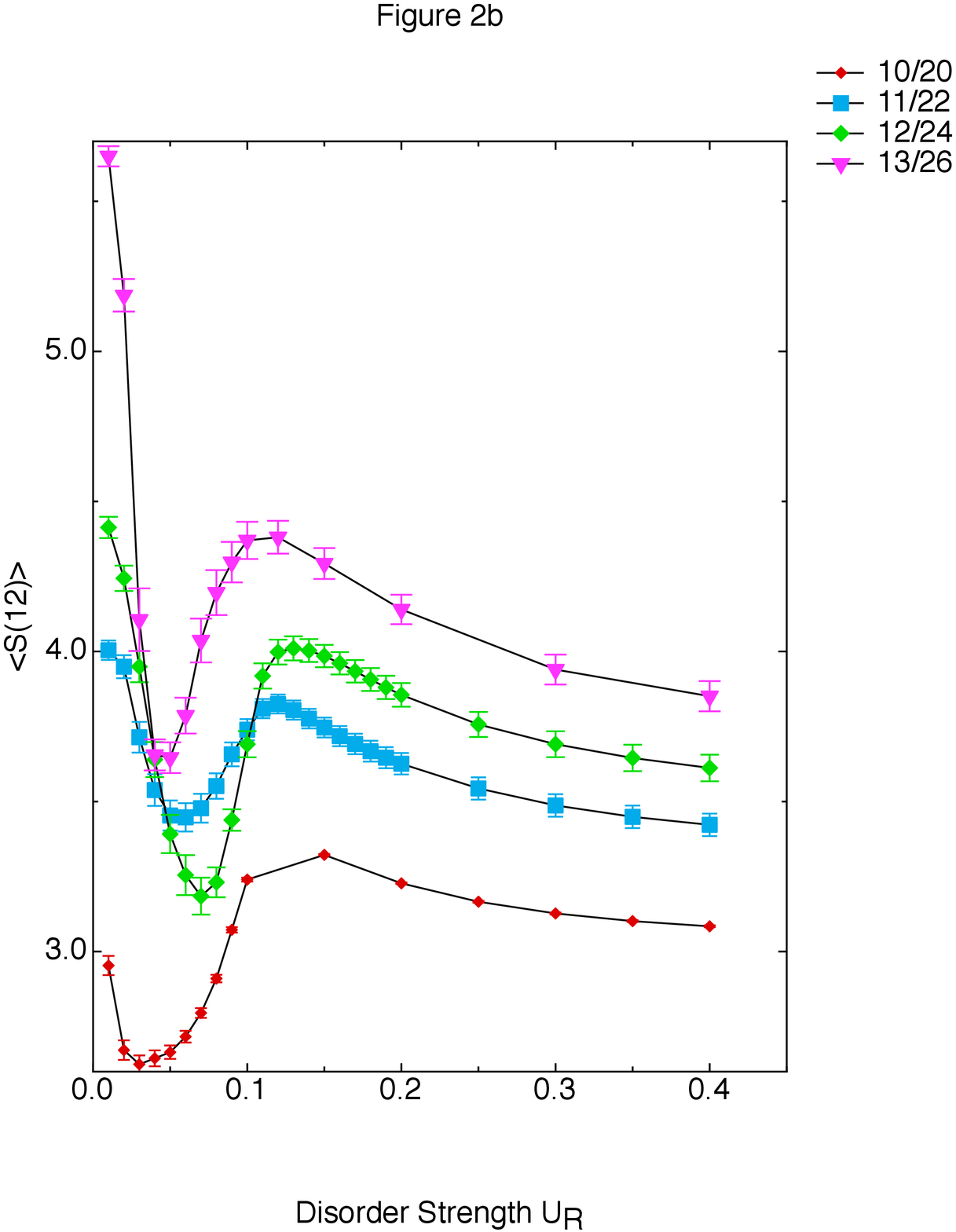}
\end{figure}

\begin{figure}[ht]
\includegraphics[height=22cm]{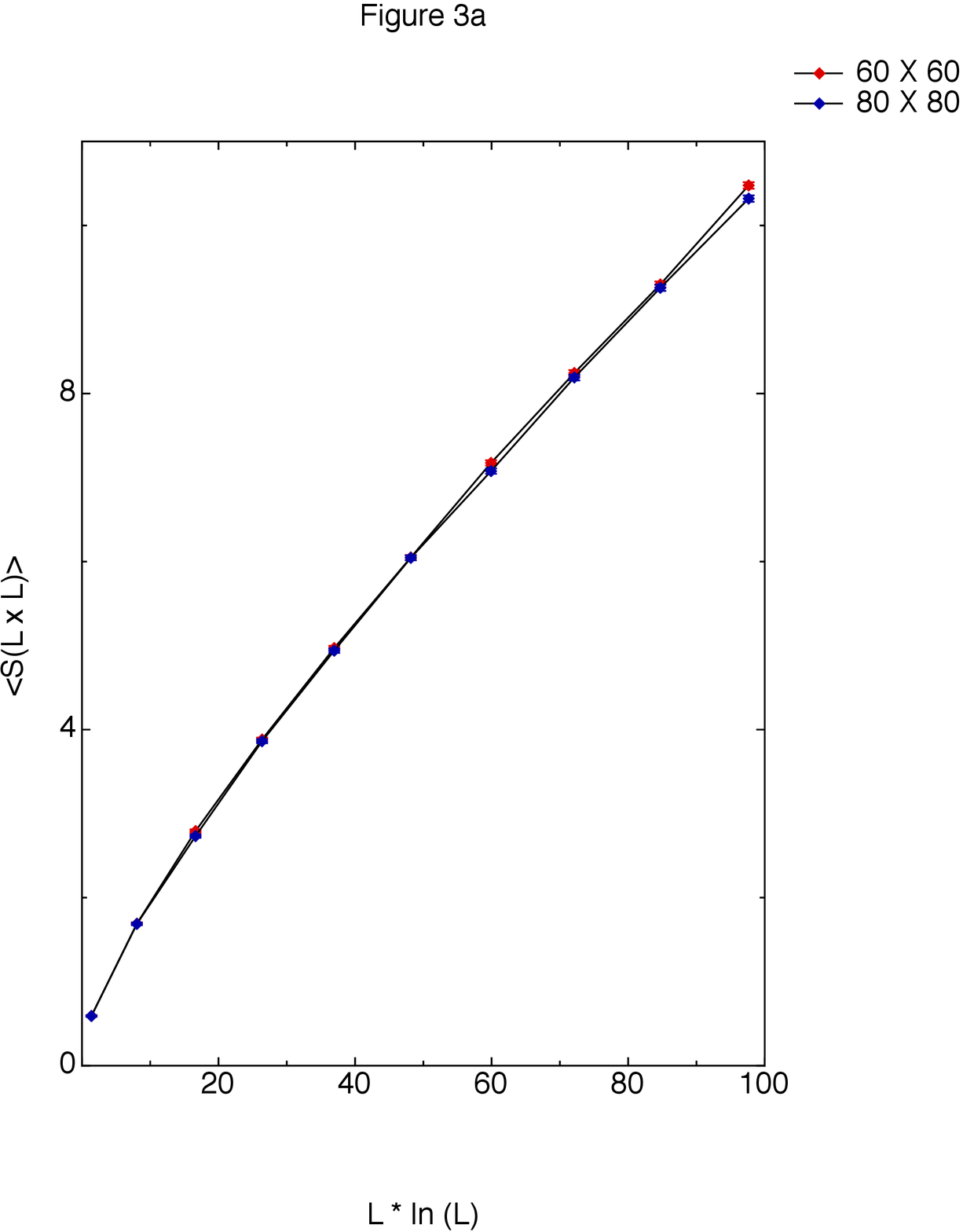}
\end{figure}

\begin{figure}[ht]
\includegraphics[height=22cm]{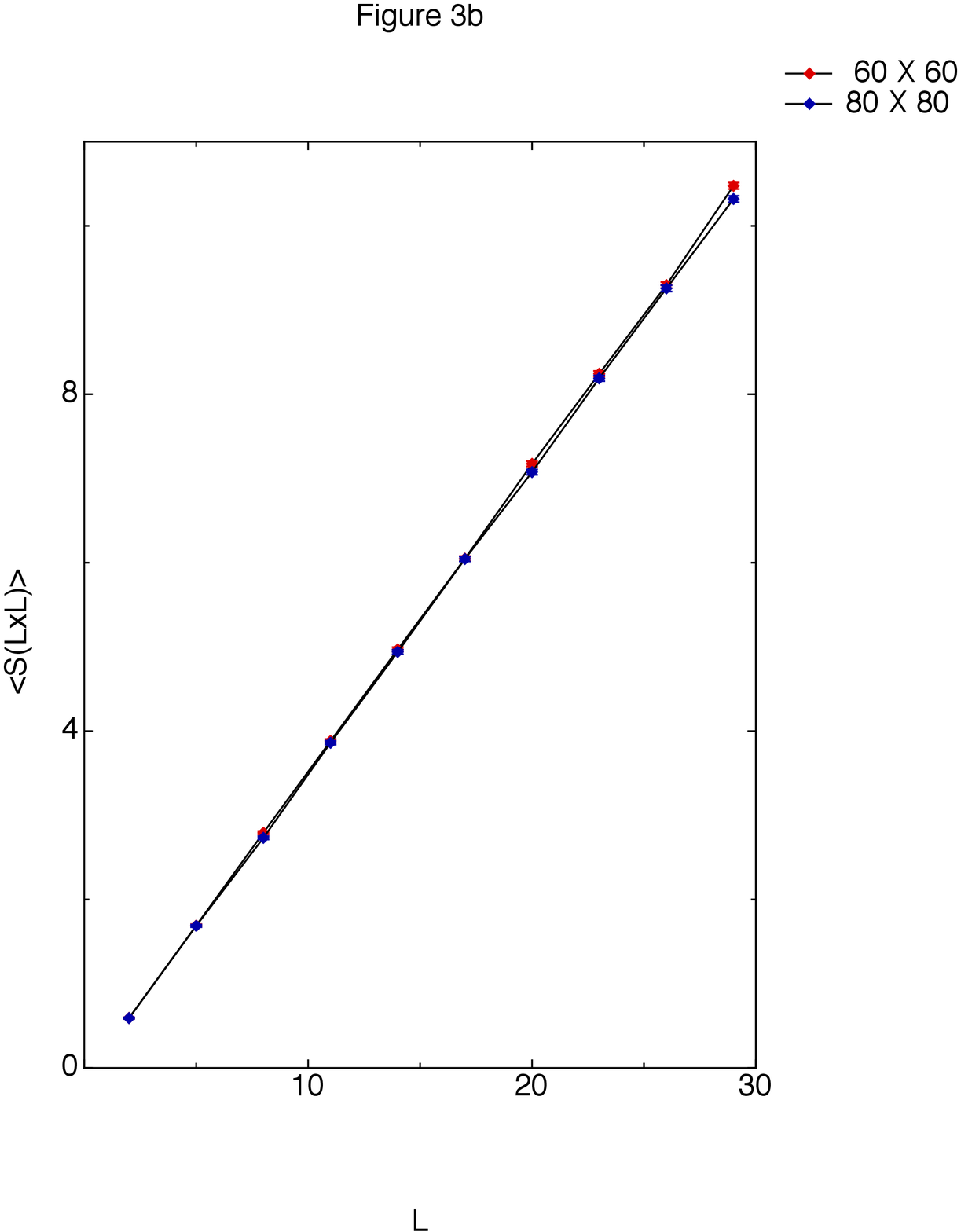}
\end{figure}

\begin{figure}[ht]
\includegraphics[height=22cm]{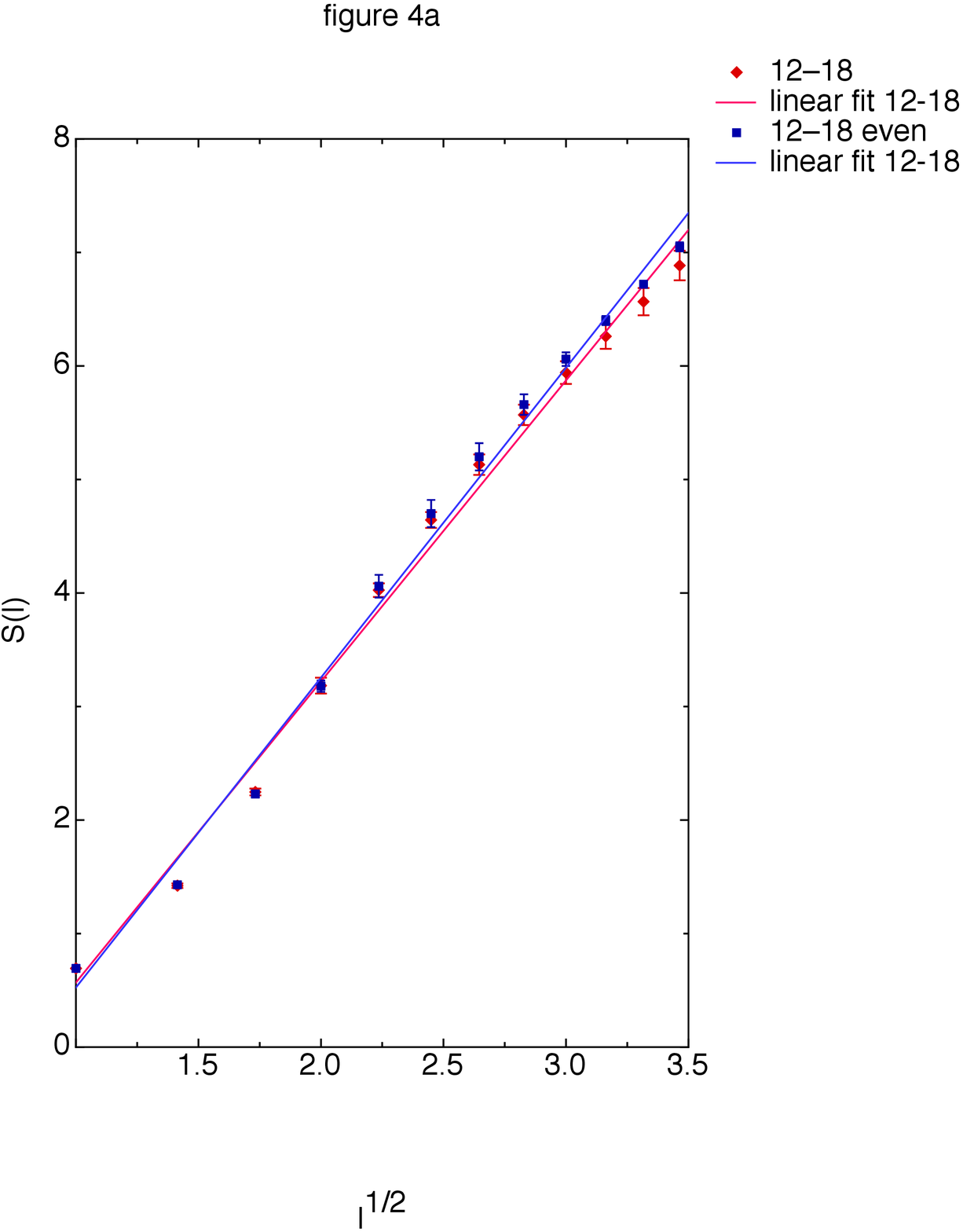}
\end{figure}

\begin{figure}[ht]
\includegraphics[height=20cm]{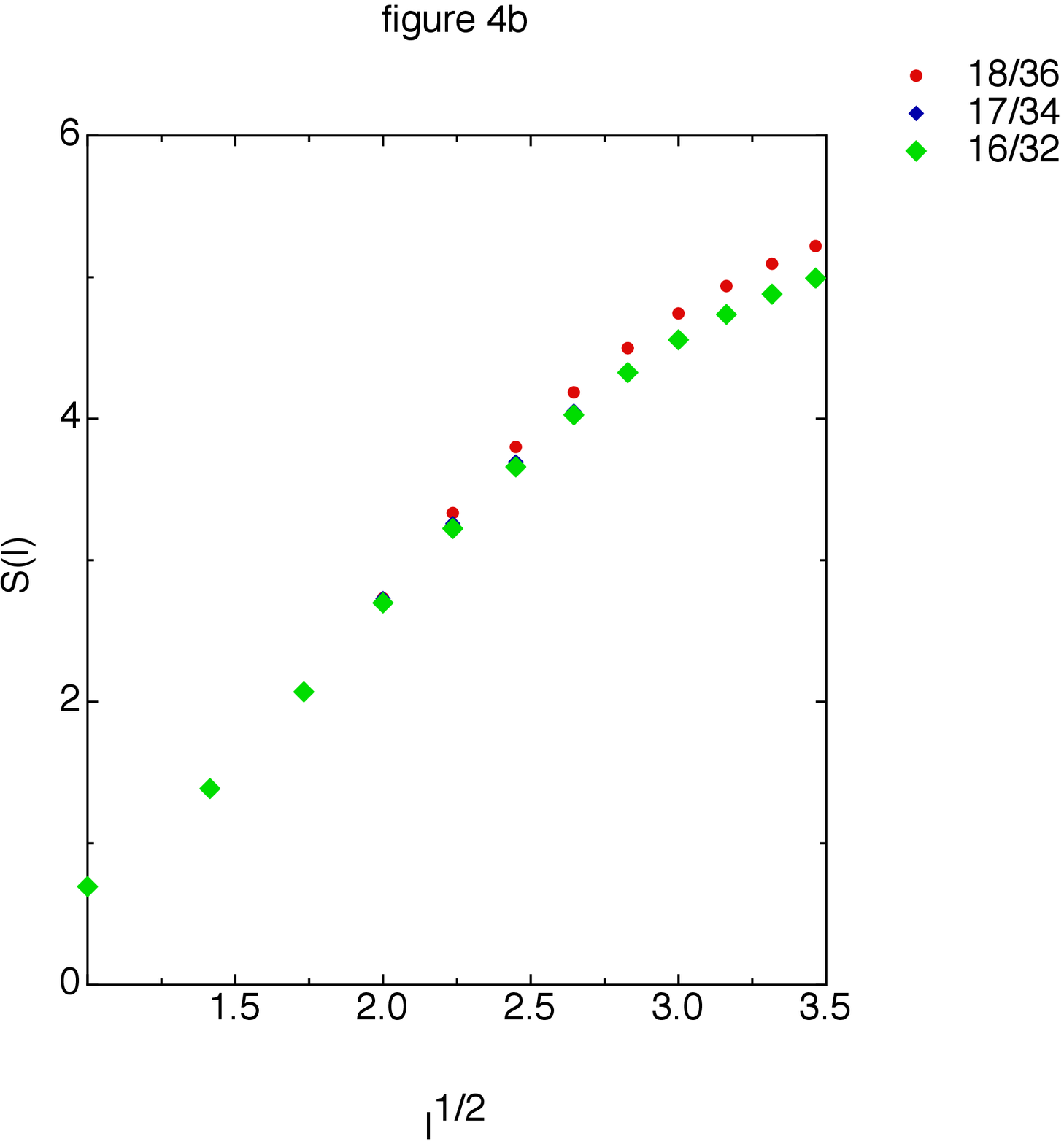}
\end{figure}

\begin{figure}[ht]
\includegraphics[height=20cm]{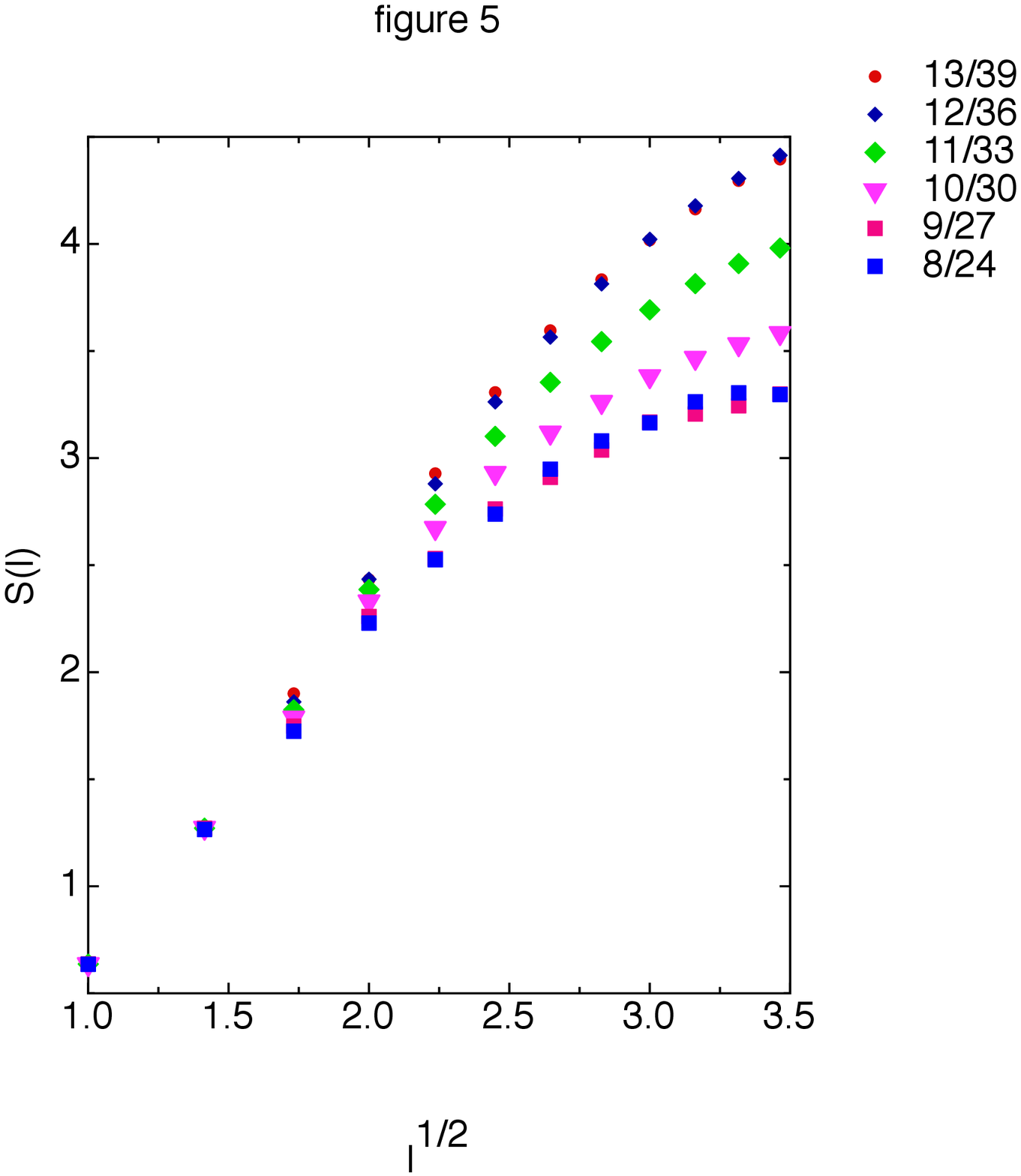}
\end{figure}

\end{document}